\documentstyle[twocolumn]{jpsj}

\def\runtitle{Dynamic Exponent of $t$-$J$ and $t$-$J$-$W$ Model} 
\def\runauthor{Hirokazu {\sc Tsunetsugu} and Masatoshi {\sc Imada}}

\title
{
Dynamic Exponent of $t$-$J$ and $t$-$J$-$W$ Model 
}

\author
{ 
Hirokazu Tsunetsugu and Masatoshi Imada$^{1}$
}

\inst
{
Institute of Applied Physics, University of Tsukuba, Tsukuba 305-8573\\
$^{1}$Institute for Solid State Physics, University of Tokyo, Tokyo 106-8666
}

\recdate
{
Mar.\ 13, 1998 
}

\abst
{
Drude weight of optical conductivity is calculated at zero 
temperature by exact diagonalization for the two-dimensional $t$-$J$ model 
with the two-particle term, $W$.  For the ordinary $t$-$J$ model with $W$=0, 
the scaling of the Drude weight 
$D \propto \delta^2$ for small doping concentration $\delta$ is obtained, 
which indicates anomalous dynamic exponent $z$=4 of the Mott transition.
When $W$ is switched 
on, the dynamic exponent recovers its conventional value $z$=2.  
This corresponds to an incoherent-to-coherent transition 
associated with the switching of the two-particle transfer.  
}

\kword
{
strongly correlated electrons, optical conductivity, Drude weight, 
Mott transition  
}

\begin{document}
\sloppy
\maketitle

High-$T_{\rm c}$ superconductors (HTSC) are a typical example of 
strongly correlated electron systems, and show many unusual 
physical properties which challenge the picture of the standard 
Fermi liquid theory.~\cite{review1,review2,review3,anderson} 
In particular, transport properties 
are among the most peculiar ones. 
Electric resistivity exhibits a linear temperature dependence 
$ \rho (T) \sim T$ near the optimal doping, and the frequency 
dependence of optical conductivity is long-tailed, 
$\sigma (\omega ) \sim \omega^{-1}$.  These are not compatible 
with the standard Fermi liquid theory, which predicts 
$\rho (T) \sim T^2$ and $\sigma (\omega ) \sim \omega^{-2}$, 
and indicate that electric transport in HTSC is anomalously 
incoherent.  Similar incoherent transport 
properties have also been observed in many other strongly 
correlated systems including a number of transition-metal 
oxides~\cite{review1} and organic conductors,~\cite{organic} 
and therefore it may be natural to ascribe the incoherent character 
to strong correlation effects.  

Concerning the mechanism of high-$T_{\rm c}$ superconductivity, 
many experiments indicate that the strong electron-electron correlation 
may play an essential role, and various theoretical approaches 
have been proposed with emphasizing different aspects of the  
correlation effects.~\cite{review1,review3,anderson}  
As discussed above, incoherent motion of electrons is a 
notable ingredient of the strong correlation effects, and 
one possible scenario is that the incoherence in the normal phase 
may drive the superconductivity so that the system can acquire 
an additional energy due to coherent motion which is recovered 
in the superconducting phase.~\cite{anderson,imada} 

This idea has recently been explored in terms of the 
quantum criticality of the Mott transition,~\cite{imada} 
where singularity of two quantities has 
been numerically investigated.  One is 
the doping dependence of compressibility $\kappa$,~\cite{furukawa,kohno}
and the other is the localization length of the single-electron 
Green's function, $\xi_l$,  
with varying chemical potential in the insulator phase.~\cite{assaad3}
Both results support the scaling hypothesis of the Mott transition with 
dynamic exponent $z$=4, larger than the conventional value $z$=2.  
If this is the case, carrier transport is expected to be 
largely incoherent.  
However, this has not been examined directly.  
It is imperative, for the direct experimental relevance of the above idea,
to clarify the incoherence of transport properties. 
Moreover the singular behavior of compressibility could allow an 
alternative interpretation, the spin density wave (SDW) 
or equivalently the nested Fermi surface.  

In this paper, we  
investigate coherence of electronic transport 
in strongly correlated electron systems 
by numerical calculations for the $t$-$J$ model.  
First, we calculate the doping dependence 
of the Drude weight at zero temperature.  The result shows the failure 
of the SDW picture and supports the scaling theory.  
We also determine the dynamic exponent $z$ to 
quantify how coherence is suppressed on approaching 
the Mott transition.  It turns out that 
the spatial dimensionality is crucial 
and the two-dimensional (2D) $t$-$J$ model is much more 
incoherent than the 1D case.  
Secondly, we explore whether the incoherent 
nature in the 2D system could be changed to evolve coherence 
by including additional electron-electron interactions.  
Recently, Assaad {\it et al.}~\cite{assaad1,assaad2,assaad4} 
found an insulator-superconductor 
transition in the 2D half-filled $t$-$U$-$W$ model, {\it i.e.}, 
the Hubbard model with additional two-electron processes.  
Supplementing the $t$-$J$ model with a similar term, we  
calculate the Drude weight and determine $z$ to see 
if coherence changes.

The Hamiltonian to investigate is the 2D $t$-$J$ model~\cite{2dtj} 
and its variant generalized by including the square of local 
kinetic energy. 
For later convenience, we include an Aharanov-Bohm (AB) 
flux, $\mbox{\boldmath $\phi$}$, and then 
the Hamiltonian using standard notation is given as 
\begin{eqnarray}
  &&{\cal H} = {\cal H}_{tJ}+{\cal H}_{W}, \label{ham}\\
  &&{\cal H}_{tJ} = - t \! \sum_{\langle i,j\rangle, \sigma} \!
  {\cal P} 
  ( \chi_{i,j} c_{i \sigma}^\dagger c_{j \sigma} \mbox{+H.c.} ) 
  {\cal P} 
  \mbox{+}  J \sum_{\langle i,j\rangle} 
   {\mib S}_{i} \cdot {\mib S}_{j} , \label{hamtj}\\ 
  &&{\cal H}_{W} = -W \sum_{j} 
  {\cal P} 
  \Bigl[ \sum_{\delta, \sigma} 
       ( \chi_{j,j+\delta} c_{j \sigma}^\dagger c_{j+\delta \sigma} 
        \mbox{+H.c.}) 
  \Bigr]^2 \!
  {\cal P}. \label{hamw} 
\end{eqnarray}
Here ${\cal P}$ is the Gutzwiller projection 
operator excluding double occupancy at every site, and 
the sum $\sum_\delta$ 
should be taken over nearest-neighbor sites. 
The phase factor is given by 
$\chi_{i,j} \equiv \exp [ -i \mbox{\boldmath $\phi$} 
 \cdot ({\mib r}_i \mbox{$-$} {\mib r}_j) / a \phi_0]$, 
where $\phi_0$ is flux quantum divided
by $2\pi$, $a$ is the lattice constant, and we shall set $a$=1,  
$\hbar$=1, and the velocity of light $c$=1 throughout this paper.  
It should be noted that $\mbox{\boldmath $\phi$}$ is constant, 
corresponding to a pure gauge, and therefore 
the magnetic field is zero everywhere.  

One may derive the main part of the $W$-term\cite{assaad1,assaad2,assaad4}  
by the strong coupling 
expansion of the Hubbard model, and then $W \propto t^2/U$ with $U$ 
being the on-site Coulomb repulsion.  It is also useful 
to rewrite the $W$-term by expanding the square in eq.~(\ref{hamw}) 
taking account 
of the projection operators.  The result is 
\begin{eqnarray}
  &&{\cal H}_{W} = 
  W \sum_{j, \sigma \sigma'} 
  \sum_{\langle \delta, \delta' \ne \delta \rangle} 
  {\cal P} 
  (2 c_{j \sigma}^\dagger c_{j \sigma'} 
   - \delta_{\sigma \sigma'} ) \nonumber\\
  && \times [ \chi_{j+\delta,j+\delta'} 
    c_{j+\delta  \sigma'}^\dagger c_{j+\delta' \sigma} 
   +\chi_{j+\delta',j+\delta} 
    c_{j+\delta' \sigma'}^\dagger c_{j+\delta  \sigma} ] 
   {\cal P}  \nonumber\\ 
  &&
  + 8W \sum_{\langle i,j \rangle } 
   ( {\mib S}_{i} \cdot {\mib S}_{j} 
  + {\textstyle {1 \over 4}} n_i n_j ) 
  -2 \zeta N_{e} W 
  , 
  \label{wterm}
\end{eqnarray}
where $\zeta$ is the number of nearest-neighbor sites 
($\zeta =4$ for a square lattice) 
and $N_e$ is the number of electrons.  
We note that as shown in eq.~({\ref{wterm}), the $W$-term 
contains so-called three-site terms~\cite{3site} and also 
renormalizes the value of $J$ in ${\cal H}_{tJ}$.  

Optical conductivity at $T$=0 may be defined 
via the Kubo formula as:~\cite{kubo} 
\begin{equation} 
  \sigma_{\mu \mu} (\omega ) 
  = {1 \over V} {i e^2 \over \omega \mbox{+} i \eta} 
  \Bigl[ 
  \langle -F_{\mu \mu} \rangle - \sum_{\alpha \ne 0} 
  { 2 (E_{\alpha} \mbox{$-$} E_{0}) 
    | \langle \alpha | J_{\mu}^P | 0 \rangle |^2 
    \over 
    (E_{\alpha} \mbox{$-$} E_{0})^2 - (\omega \mbox{$+$}  i \eta)^2 } 
  \Bigr], 
  \label{kubo}
\end{equation} 
where $-e$ is the electron charge, $\mu$ denotes the spatial index, 
$V$ is the number of sites, and $\eta$ is an adiabatic constant.  
$|\alpha\rangle$'s are eigenstates of the Hamiltonian with 
eigenenergy $E_\alpha$, where $|0\rangle$ 
denotes the ground state, and 
\begin{equation}
  J_\mu^P = {1 \over e}
  {\partial {\cal H} \over \partial \phi_{\mu}} , 
  \ \ \ 
  F_{\mu \mu'} =
  - {1 \over e^2} 
  {\partial^2 {\cal H} \over \partial \phi_{\mu} \partial \phi_{\mu '}} .  
  \label{JF}  
\end{equation}
We may separate the Drude part from the formula~(\ref{kubo}) 
and the remaining is the regular part: 
\begin{equation}
  \label{split}
  \sigma_{\mu \mu} (\omega) = 
  { \pi e^2 \over V} D_{\mu} \delta( \omega ) 
  + \sigma_{\mu \mu}^{\rm reg} (\omega )  .  
\end{equation}
The Drude weight is simply given by~\cite{kubo} 
\begin{equation}
  D_{\mu} = 
  \langle -F_{\mu \mu} \rangle - 2 \sum_{\alpha \ne 0} 
  { | \langle \alpha | J_{\mu}^P | 0 \rangle |^2 
    \over 
    E_{\alpha} \mbox{$-$} E_{0} } 
  = {1 \over e^2}
  {\partial^2 E_0 (\mbox{\boldmath $\phi$})
   \over 
   \partial \phi_{\mu}^2 } . 
  \label{drude}
\end{equation}

In addition to the Drude weight, another important quantity 
is the total weight of the optical conductivity.  
It is straightforward to rewrite the Kubo formula~(\ref{kubo}) 
and obtain the following sum rule~\cite{kubo} 
\begin{equation}
  \int_{-\infty}^{\infty} \sigma_{\mu \mu} (\omega ) {\rm d}\omega 
  = {\pi e^2 \over V} 
  \langle -F_{\mu \mu} \rangle  .
  \label{sum}
\end{equation} 
It is noted that if $W$=0, $\langle F_{\mu \mu} \rangle$ reduces to 
the kinetic energy along the $\mu$-direction, while 
it does not 
if $W$$\ne$0 or if long-range electron hoppings are introduced.  

To calculate the Drude weight numerically, we have first obtained 
the ground state of ${\cal H}$ by the Lanczos 
method, and then employed the first relation in eq.~(\ref{drude}).   
For the latter step, we have used the Lanczos method again.  
The result agrees well with the value calculated via 
the second relation in eq.~(\ref{drude}), and the relative error 
is typically as small as $10^{-6}$.  The lattice used in this 
work is a 4$\times$4-site cluster with periodic boundary conditions 
in both directions, 
$\mbox{\boldmath $\phi$}$=$(0,0)$ and the total wave number 
${\mib K}=(0,0)$, unless explicitly mentioned.  
Due to the symmetry of the cluster, the conductivity is isotropic: 
$\sigma_{\mu \mu} = \sigma$, $D_{\mu}=D$, 
and $\langle-F_{\mu \mu'}\rangle =\langle- F\rangle \delta_{\mu \mu '}$.  

First we show the results for the 4$\times$4-site $t$-$J$ model 
(i.e., $W$=0) at various $J/t$'s.  
The Drude weight $D$ and the total weight $\langle -F \rangle$ 
are plotted as a function of electron density $n_e \equiv N_e /V$ 
in Fig.~\ref{fig1}(a).   
The total weight $\langle -F \rangle$ has a maximum at 
$n_e$$\approx$0.5, and it is nearly symmetric around it.  
In contrast, the Drude weight is considerably reduced 
from $\langle -F \rangle$ in the region of $n_e$$>$0.5.  

Similar calculations were performed for a 2D $t$-$J$ model 
by Dagotto {\it et al.}~\cite{dagotto} to study the dependence on 
hole doping $\delta$=$\mbox{1$-$$n_e$}$, 
and they concluded $D \sim \delta$ as $\delta \rightarrow 0$.  
We have reexamined the doping dependence and found a different scaling.  
To this end, it is convenient to calculate first the 
$\delta$-dependence of the ratio of $D$ 
to $\langle -F \rangle$ 
and then transform the result to that of $D$.   
Fig.~\ref{fig1}(b) shows $D/\langle -F \rangle$ 
calculated from the data in Fig.~\ref{fig1}(a),  
as a function of electron density.  
It is clear that this ratio becomes smaller with 
decreasing hole doping, indicating 
$D/\langle -F \rangle \sim \delta$ as $\delta \rightarrow 0$.  
The kinetic energy $\langle -F \rangle$ contains not only 
the coherent part of electron motion but also the incoherent 
part. Therefore it is natural that 
$\langle -F \rangle \sim \delta$, and the data shown in 
Fig.~\ref{fig1}(a) confirms this scaling.  Combining these 
$\delta$-dependences, we obtain 
$D \sim \delta^2$.  

The result for the 1D $t$-$J$ model is very different from 
the 2D case.~\cite{stephan}  We plot in Fig.~\ref{fig2} the electron-density 
dependence of $D$ and $\langle -F \rangle$ for various system 
sizes at $J/t$=0.3.  
Difference between the two weights, i.e.\ the integrated weight of 
$\sigma^{\rm reg} (\omega)$, is nearly zero at this scale.  
Therefore, not only $\langle -F \rangle \sim \delta$, but 
also $D \sim \delta$ as $\delta \rightarrow 0$ in the 1D case.  
This result is consistent with the Luttinger liquid theory.~\cite{schulz} 
It predicts $D = 2 v_c K_c$, where 
$v_c$ is the charge velocity and asymptotically $v_c \sim \delta$.  
$K_c$ is the Tomonaga-Luttinger liquid 
parameter, and $K_c \approx {1 \over 2}$
in the strong correlation limit, leading to $D \sim \delta$.  

The different $\delta$-dependence of $D$ between 1D and 2D cases 
results in 
different values of the dynamic exponent $z$.  The dynamic exponent 
$z$ is defined by the divergence of correlation time $\tau$ and 
correlation length $\xi$ on approaching the critical point: 
$\tau \sim \xi^z$.  Scaling theory for the Mott transition 
by Imada~\cite{imada} predicts $ D \sim \delta^{1+(z-2)/d}$, 
where $d$ is the spatial dimension.  As shown above, 
$ D \sim \delta^2$ and $\delta^1$ 
for 2D and 1D $t$-$J$ model, respectively, and this implies 
$z$=4 for the 2D and $z$=2 for the 1D cases.  The dynamic exponent 
is a measure for characterizing the coherence of the system: the larger 
it is, the faster the coherent part of conductivity vanishes at 
the Mott transition, indicating that the system 
is more incoherent.  
Thus, coherence is very sensitive to the spatial dimensionality 
of the system, and the 2D $t$-$J$ model is more incoherent 
than the 1D model. 

The present result for $z$ is consistent with other numerical studies, 
if one is based on the scaling theory.  
One is the compressibility of the 2D Hubbard model,~\cite{furukawa} 
where $\kappa \sim \delta^{-1}$ 
near half filling.  A similar behavior was also found for 
the 2D $t$-$J$ model by Kohno.~\cite{kohno}  
According to the scaling theory, these results 
indicate $z$=4 
and agree with the present study.  
Another support is the chemical-potential dependence 
of the localization length $\xi_l$.~\cite{assaad3}  It is found that 
$\xi_l \sim |\mu - \mu_c|^{-\nu}$ with $\nu$=${1 \over 4}$, 
and combining with the relation $\nu z$=1, this also gives $z$=4.
An alternative explanation for the $\delta$-dependence of $\kappa$ 
may be given by the SDW picture, which 
gives the energy $E (\delta) \! \sim \! \Delta \delta \mbox{+} 
A \delta^3$ in the 2D case, with $\Delta$ being the momentum-independent 
quasiparticle gap.   This implies 
$\kappa = [(\mbox{1$-$$\delta$})^2 E''(\delta) ]^{-1} \sim \delta^{-1}$, 
consistent 
with the numerical results.  In the SDW picture, however, 
the kinetic energy is carried by {\it coherent} quasiparticles,  
and therefore both the Drude weight and total weight should have 
a $\delta^1$-term, which is not the case 
as shown in Fig.~\ref{fig1}.  

Next, we study the stability of the incoherent phase with $z$=4  
found for the 2D $t$-$J$ model and study what type of 
extra processes could drive the evolution of the coherence.  A similar idea 
has been explored by Assaad {\it et al.} for the 
2D Hubbard model,~\cite{assaad1,assaad2,assaad4} 
where 
${\cal H}_W$ of eq.~(\ref{wterm}) without the projection 
operator ${\cal P}$ was introduced.  

We show the result for the 2D $t$-$J$-$W$ model in Fig.~\ref{fig3}(a).
The electron-density dependence 
of the Drude weight and total weight is plotted 
for $J/t$=0.3 and various values of $W$.  
For $N_e$=14 and 12, the ground state is obtained 
for the AB flux 
$\mbox{\boldmath{$\phi$}}$=$({\pi \over 4},{\pi \over 4})$, 
and this point will be discussed later.  
With increasing $W$, the Drude weight 
and the total weight both increase.  This is because 
the $W$-term introduces additional processes of 
transfer involving eight second-neighbor sites,  
which yields a gain of the corresponding 
energy described in eq.~(\ref{wterm}).  

An important effect of the $W$-term is that it affects 
the ratio $D/\langle -F \rangle$ noticeably as shown in 
Fig.~\ref{fig3}(b).  
In particular, the ratio 
increases drastically near half-filling, 
and seemingly converges to a finite 
value as $\delta \rightarrow 0$.  
This is in contrast to the case for $W$=0, where the ratio 
vanishes linearly with $\delta$.  
If this is the case, it means that the scaling of 
the Drude weight is now $D \sim \delta^1$ as in the 1D $t$-$J$ model, 
distinct from $D \sim \delta^2$ at $W$=0.  
Therefore, the dynamic exponent is correspondingly modified to $z$=2, 
the conventional value as in uncorrelated electron systems, 
and we can infer that the system is stabilized by gaining an energy 
of coherent motion driven by the $W$-term.  
A similar change in $z$ has also been observed for the 2D 
$t$-$U$-$W$ model,~\cite{assaad4} where the exponent for 
the single-electron localization length changes from 
$\nu$=${1 \over 4}$ to ${1 \over 2}$.  

It is interesting that the Drude weight calculated at 
$\mbox{\boldmath{$\phi$}}$=(0,0) is negative at $N_e$=14 
for $W/t \ge 0.3$.  This means that a finite AB flux 
is induced spontaneously.  
To see this in more detail, we have calculated the 
$\mbox{\boldmath{$\phi$}}$-dependence of the lowest eigenenergy 
for all the sixteen wave vectors ${\mib K}$.  For the 
system size of 4x4 sites, the system with 
$\mbox{\boldmath{$\phi$}}$=$({p\pi \over 2},{q\pi \over 2})$ 
($p$,$q$: integer and, $\mbox{\boldmath{$\phi$}}$ is in units of $\phi_0$)
has the same energy spectrum as that with 
$\mbox{\boldmath{$\phi$}}$=$(0,0)$.  Therefore it is sufficient 
to vary $\mbox{\boldmath{$\phi$}}$ in a trianglular region 
defined by the three points, $(0,0)$, $({\pi \over 4},0)$, and 
$({\pi \over 4},{\pi \over 4})$.  We have kept track of the lowest 
eigenenergy for all the ${\mib K}$'s along the edges of the trianglular 
$\mbox{\boldmath{$\phi$}}$-region, 
and the result is shown in Fig.~\ref{fig4}.  
The eigenstates have the lowest energy at 
$\mbox{\boldmath{$\phi$}}$=$({\pi \over 4},{\pi \over 4})$ and 
${\mib K}$=$(-{\pi \over 2},-{\pi \over 2})$, when $W/t \ge 0.05$.  
A finite AB flux is also found for $N_e$=12 when $W/t \ge 0.1$, and 
the value is again 
$\mbox{\boldmath{$\phi$}}$=$({\pi \over 4},{\pi \over 4})$, 
but now at ${\mib K}$=$(\pi,\pi)$.  

This spontaneous AB flux may be an evidence of enhancement 
of $d_{x^2-y^2}$-wave superconducting correlation due to the $W$-term.  
Actually, such a superconducting phase has been observed 
in the $t$-$U$-$W$ model.~\cite{assaad2,assaad4}
Recall that the finite flux shifts one-particle momentum as 
${\mib k}$$\rightarrow$${\mib k}$$-$$\mbox{\boldmath{$\phi$}}$, 
and no electrons then have the renormalized momentum 
${\mib k}'=(\pm {\pi \over 2},\pm {\pi \over 2})$.  
Therefore all electrons on the tight-binding ``Fermi surface'' 
can utilize a pairing potential with $d_{x^2-y^2}$ symmetry.
This also suggests the importance of strong momentum dependence of
charge excitations, and that incoherence 
around $(\pm \pi,0)$ and $(0,\pm \pi)$
rather than $(\pm {\pi \over 2},\pm {\pi \over 2})$ 
is responsible for the retrieval of 
the coherence and the pairing. We will discuss this point 
in more detail elsewhere.

In this paper, we have numerically studied coherence character 
in transport properties of strongly correlated electron systems.   
We have calculated the doping dependence 
of the Drude and total weights of optical conductivity 
for the 2D and 1D $t$-$J$ model at zero temperature.  
The result shows different scalings near half filling, 
$D \sim \delta^{2}$ and $\delta^{1}$ for the 2D and 1D case, 
respectively.  The dynamic exponent of the Mott transition as 
$\delta \rightarrow 0$ is determined as 
$z$=4 for the 2D case while $z$=2 for the 1D case.  
Incoherent transport properties in the 2D system may 
be attributed to this anomalous dynamic exponent.  
We have also examined the effects of the extra electron-electron 
interaction, ${\cal H}_W$, in two dimensions.  It turns out that 
the dynamic exponent changes to $z$=2 and coherence is retrieved.  

The authors thank Fakher Assaad for fruitful discussions.  
This work was supported by JSPS "Research for the Future" Program 
(JSPS-RFTF97P01103)
as well as by a Grant-in-Aid for Scientific Research 
from the Ministry of Education, Science, Sports, and Culture 
of Japan.  Numerical calculations were mainly performed 
on the VPP500 at the Institute for Solid State Physics, 
University of Tokyo, 
and the VPP500 at the Science Information Processing Center, 
University of Tsukuba.

\begin{figure}[hbt]
\caption{(a) Drude and total weights of the optical conductivity 
of the 2D $t$-$J$ model with 4$\times$4 sites at $T$=0.
(b) The ratio of the two weights $D/\langle -F \rangle$.}  
\label{fig1}
\end{figure}

\begin{figure}[hbt]
\caption{Drude and total weights of the optical conductivity 
of the 1D $t$-$J$ model at $T$=0 for $J/t$=0.3. }
\label{fig2}
\end{figure}

\begin{figure}[hbt]
\caption{(a) Drude and total weights of the optical conductivity 
of the 2D $t$-$J$-$W$ model with 4$\times$4 sites at $T$=0 for $J/t$=0.3.  
$D$: open symbols, and $\langle -F \rangle$: solid symbols.  
For $n_e$=${14 \over 16}$ and ${12 \over 16}$, 
the ground state is calculated at 
$\mbox{\boldmath $\phi$}$=$({\pi \over 4},{\pi \over 4})$.   
(b) The ratio of the two weights $D/\langle -F \rangle$.  
(c) The $W$-dependence of the ratio $D/\langle -F \rangle$ for
$n_e$=${14 \over 16}$ at $J/t$=0.3. }
\label{fig3}
\end{figure}

\begin{figure}[hbt]
\caption{The AB-flux dependence of the lowest eigenenergy for all 
wave vectors for the 2D $t$-$J$-$W$ model, 
14 electrons in 4$\times$4 sites, and $J/t$=0.3, $W/t$=1.0.  
}
\label{fig4}
\end{figure}

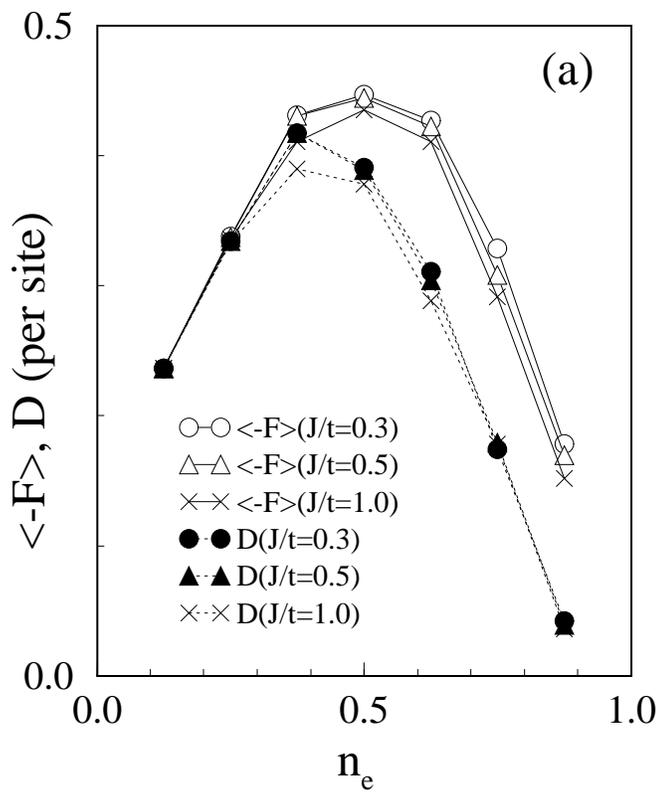
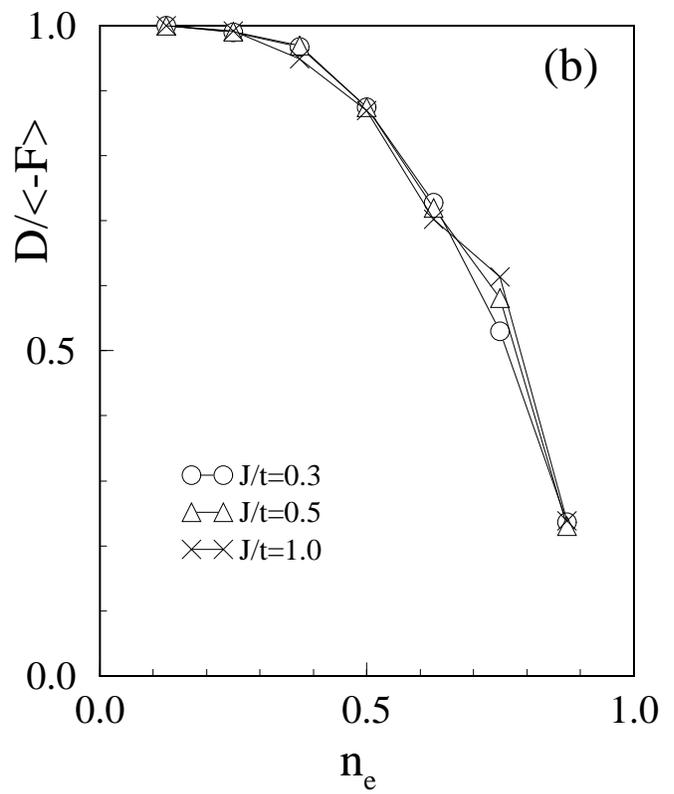
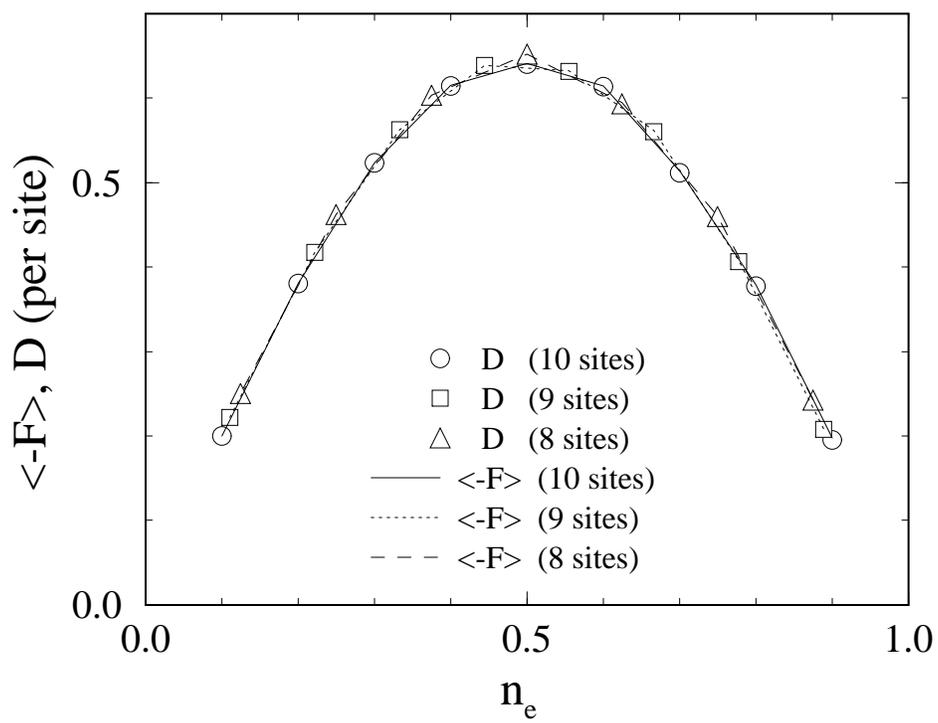
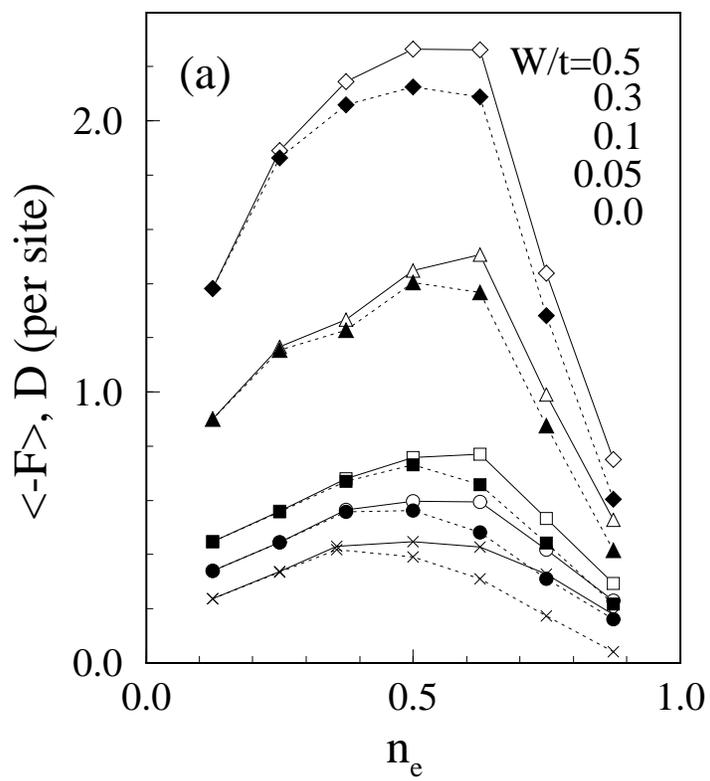
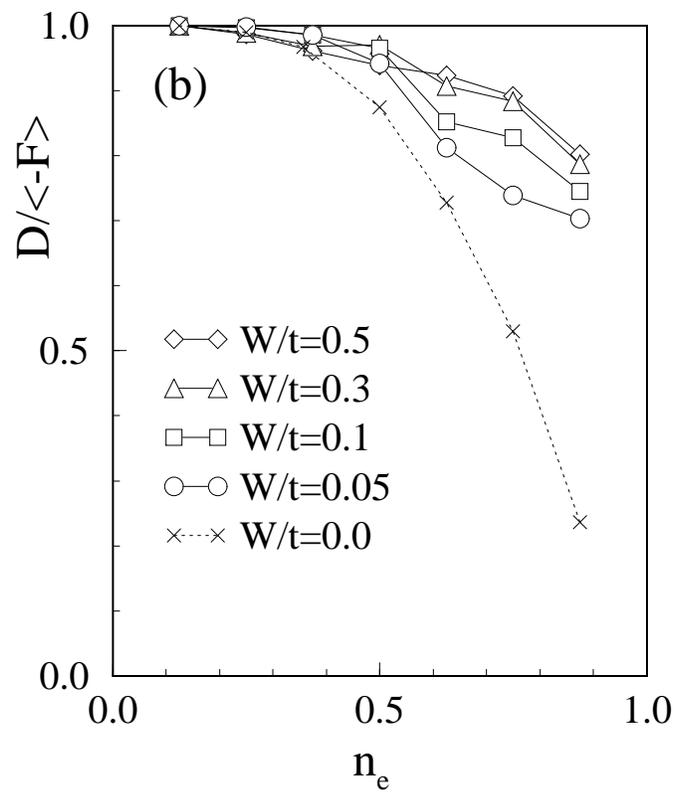
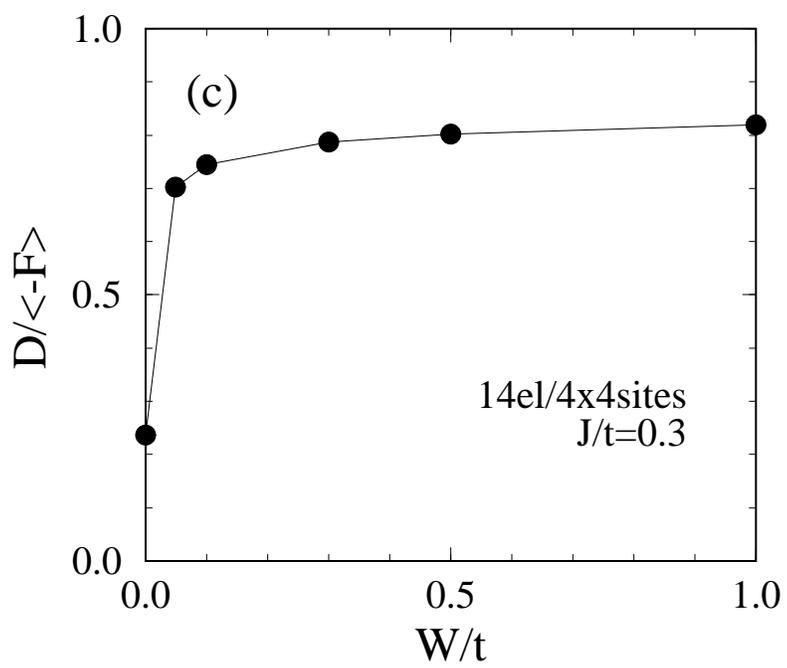
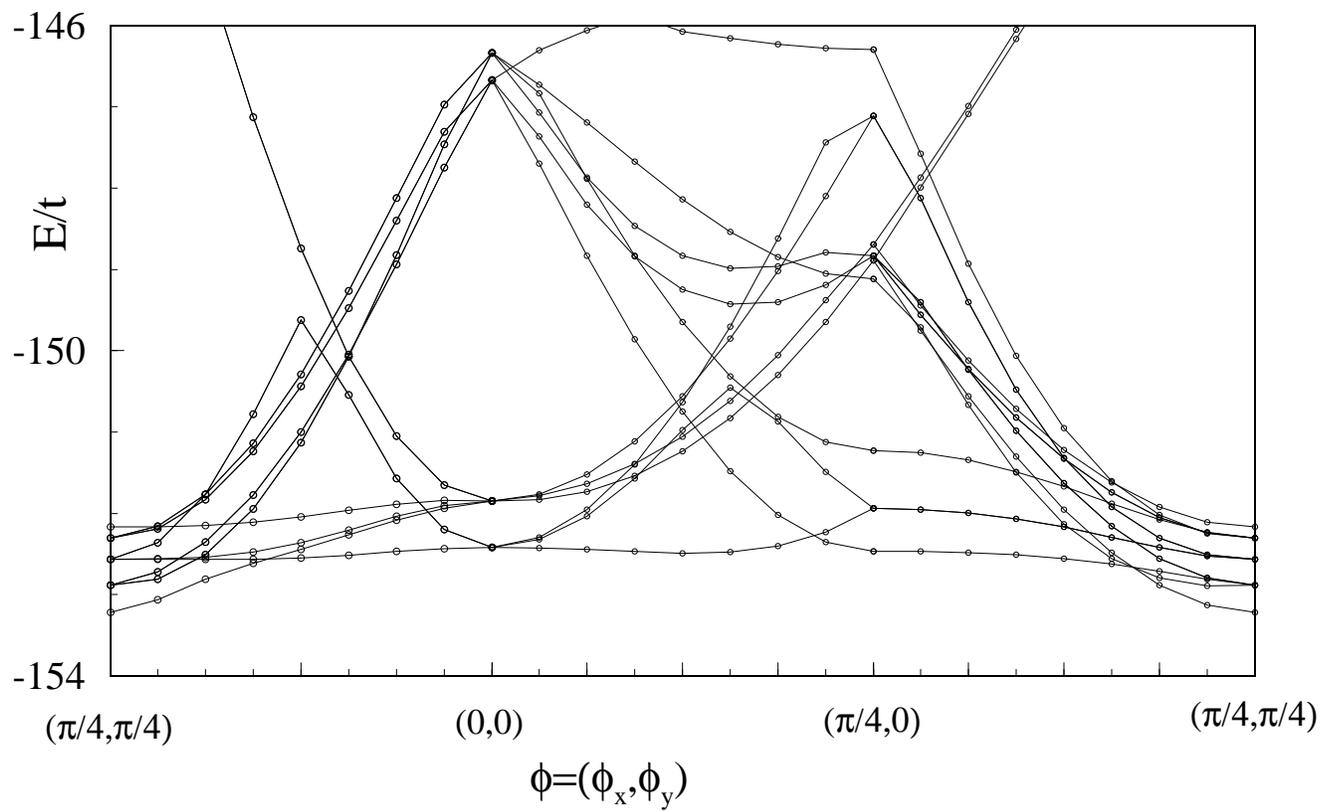
\end{document}